\documentstyle [12pt]{article}
\advance\voffset by -1 in
\advance\hoffset by -.5 in
\def\a{\alpha}
\def\b{\beta}
\def\g{\gamma}

\def\p{\partial}

\def\O{\Omega}
\def\Om0{\Omega^0}
\def\O1{\Omega^1}

\def\tp{\tilde{\p}}
\def\rarr{\rightarrow}

\def\hw{\hat{w}}

\def\res{{\rm res}}

\def\bH{{\bf H}}
\def\G{{\bf G}}
\begin{document}

\centerline{{\bf On a generalization of the Fay-Sato identity }}
\centerline{{\bf for KP Baker functions and}}
\centerline{{\bf its application to constrained hierarchies}}

\vspace{.1in}
\begin{center}
LEONID DICKEY\\

Math. Department, University of Oklahoma\\
Norman, OK 73019, USA\\
e-mail: ldickey@uoknor.edu\\

and\\

WALTER STRAMPP\\

Fachbereich 17-Mathematik/Informatik\\
Universit\"at-GH Kassel\\
Holl\"andische Strasse 36\\
34109 Kassel, Germany\\
e-mail: strampp@hrz.uni-kassel.de
\end{center}

\vspace{.3in}
\begin{abstract}Some new formulas for the KP hierarchy are derived from the 
differential Fay identity. They proved to be useful for the $k$-constrained 
hierarchies providing a series of determinant identities for them. A 
differential equation is introduced which is called ``universal" since it 
plays an important role for all the $k$-constrained hierarchies.  
In the cases $k=1,2$ and 3 explicit formulas are presented, in all the others
recurrence relations are given which enable one to obtain the identities.
\end{abstract}
\newpage
{\bf 0. Introduction.} This paper is a result of our discussions 
about the meaning and the significance of a proposition in the article [1]
(Proposition 2). We found out that a broader problem can be posed and 
solved using a more general language. Virtually, we suggest a new type of
problems for the KP and, especially, $k$-constrained KP hierarchies. 
The explicit identities are obtained for the smallest values of $k$, 
$k=1,2$ and 3; for higher values we have recurrence formulas, i.e. a
recipe to get new identities. Calculations become cumbersome. We hope that 
there exists a more general approach allowing one to circumvent these 
complicated calculations at all.

The so-called differential Fay identity for the KP hierarchy (proven by Sato 
for the most general $\tau$ functions) is well-known 
(see, e.g., [3]). It can be written in the following form using the Baker 
function $\hw(t,z_1)\exp\sum t_iz_1^i$. Let $\G$ be a shift operator defined 
as $$\G F(t_1,t_2,...)=F(t_1+1/z_2,t_2+1/(2z_2^2), t_3+1/(3z_2^3),...)$$ where
$z_2$ is another parameter and let $\G_1=\G-1$. Then $$(z_1-z_2)\G_1\hw=-\G_1
\hw'+
\G_1w_1\cdot\G_1\hw+\G_1w_1\cdot\hw-\hw'.\eqno{(0.1)}$$Here and further $\hw$ 
is always $\hw(t,z_1)$ and $w_1$ is the first non-trivial
coefficient in the expansion $\hw=\sum_0^\infty w_iz_1^{-i}$ (recall that
$w_0=1$). Let $$\bH=-\p+\G_1w_1+\hw\cdot\res_{z_1}\eqno{(0.2)}$$ be an 
operator that acts on series $\sum a_{ij}z_1^{-i}z_2^{-j}$. Eq.(0.1) can be 
represented as $$(z_1-z_2)n^{(1)}=\bH(n^{(1)})-\hw',~~{\rm where}~n^{(1)}=
\G_1\hw.\eqno{(0.3)}$$ 
In fact, this identity is equivalent to the hierarchy itself since,
expanding it in powers of $z_1^{-1}$ and $z_2^{-1}$, all the $\p_jw_i$ can be
obtained.

Let us think on the last term of (0.3) as on $-\p_1\hw$. 
We are going to generalize this identity by showing that 
one can construct quantities $n^{(k)}=\sum_1^\infty n^{(k)}_{j}z_2^{-j}$ where 
$n^{(k)}_{j}=P_j^{(k)}\hw$ and $P_j^{(k)}$ are differential operators in $\p_l$ 
such that 
$$ (z_1-z_2)n^{(k)}=\bH(n^{(k)})-\p_k\hw,~~\p_k=\p/\p t_k.\eqno{(0.4)}$$
Explicit formulas for $n^{(k)}$ are obtained only for $k=1,2,3$. In the case of 
larger $k$ we give the recurrence formulas for $n^{(k)}_{j}$.   

These identities have an interesting application in the theory of 
constrained KP hierarchies. The $k$-constraint for the KP hierarchy 
is $L^k=L_+^k+q\p^{-1}r$. An important role for all the $k$-constraints plays
an equation which we call ``universal". This is $$f'+1-z_2f+Qf
-\G_1w_1\cdot f=0 \eqno{(0.5)}$$ where $Q$ is any function. The universal 
equation has the only solution in the form of a series in $z_2^{-1}$,
$f_Q=\sum_1^\infty f_jz_2^{-j}$. The coefficients $f_j$ are functions, one
can find that $f_Q=z_2^{-1}+Qz_2^{-2}+...$. It will be shown that for 
the $k$-constrained hierarchy the quantities $n^{(k)}/\p_k\hw$ with the above
$n^{(k)}$ satisfy the universal equation with $Q=q'/q$. Therefore $n^{(k)}/\p_k
\hw=f_Q$ and they do not depend on $z_1$. The quantity $n^{(k)}$ happens to be
a product of a series in $z_1^{-1}$ and a series in $z_2^{-1}$. Hence, the 
relations $P_j^{(k)}w_i=f_j\cdot\p_kw_i$, $j\geq k$ hold where $P_j^{(k)}$ are 
the above differential operators. This implies that the matrix $P_j^{(k)}w_i$ 
is of rank 1 and all of its second order minors vanish. This supplies one 
with infinitely many identities. The first of them, with $j_1=1$ and 
$j_2=2$ and arbitrary $i_1$ and $i_2$, form a necessary and sufficient
condition that a Baker function of the KP hierarchy satisfies the 
$k$th constraint.\\ 

{\bf 1. The KP hierarchy.} Let us recall some terminology.

Let$$L=\p+u_1\p^{-1}+u_2\p^{-2}+...$$ be a
pseudodifferential operator, then $$\p_nL=[L_+^n,L],~{\rm where}~\p_n=\p/\p
t_n$$ are the equations of the KP hierarchy, $$L=\hw(t,\p)\p\hw^{-1}(t,\p),~\hw
(t,\p)=\sum_0^\infty w_k(t)\p^{-k},~w_0=1,$$and $$\psi(t,z)=\hw(t,z)\exp\xi=
\hw(t,\p)\exp\xi~{\rm where}~\xi(t,z)=\sum_1^\infty t_kz^k$$ is the (formal) 
Baker function. 

It is easy to see that the equations $$\p_n\hw(t,\p)=-L_-^n\hw(t,\p)$$ imply 
the above hierarchy equations for $L$. Then $\psi$ satisfies
the equations $$L\psi=z\psi,~~\p_n\psi=L_+^n\psi.$$
The Schur polynomials are defined by $$\exp\xi=\sum_0^\infty p_n(t)z^n,~
t=(t_1,t_2,...).$$Let $\tp=(\p_1,\p_2/2,\p_3/3,...)$.
Then $p_n(\tp)$ are differential operators denoted by $p_n^+$. \\

{\bf 2. The $\tau$-function and the differential Fay identity.}
The $\tau$-function is defined by the equality
$$\hw(t,z)={\tau(t-[z^{-1}])\over \tau
(t)}\eqno{(2.1)}$$ where $\tau(t)=\tau(t_1,t_2,...)$ and
$\tau(t-[z^{-1}])=\tau(t_1-
1/z,t_2-1/(2z^2),t_3-1/(3z^3),...)$. The existence of such a function is
not obvious and must be proven. An important step in the proof is establishment 
of a relation $${\hw(t-[z_2^{-1}],z_1)\over \hw(t,z_1)}={\hw(t-[z_1^{-1}],z_2)
\over \hw(t,z_2)}\eqno{(2.2)}$$which, in its turn, easily follows from (2.1). 
We need further a corollary of this equality that can be obtained expanding it 
in powers of $z_1^{-1}$: $${\hw'(t,z_2)\over\hw(t,z_2)}=-w_1(t-[z_2^{-1}])+w_1
(t).\eqno{(2.3)}$$ The relation (2.2) states that
$\hw(t-[z_2^{-1}],z_1)/\hw(t,z_1)$ is symmetric with respect to
$z_1$ and $z_2$ and $(z_2-z_1)\hw(t-[z_2^{-1}],z_1)/\hw(t,z_1)$ is skew
symmetric. We can do a more precise statement.\\

{\bf Lemma.} {\sl A relation 
$$(z_2-z_1){\hw(t-[z_2^{-1}],z_1)\over \hw(t,z_1)}=-{\hw'(t,z_1)\over
\hw(t,z_1)}+{\hw'(t,z_2)\over\hw(t,z_2)}+z_2-z_1$$ 
holds.}\\

{\em Proof.} As it is easy to check, this is nothing but the differential Fay 
identity (see, e.g., [3]) $$\p\tau(t-[z_1^{-1}])\tau(t-[z_2^{-1}])-\tau(t
-[z_1^{-1}])\p\tau(t-[z_2^{-1}])$$ $$=(z_2-z_1)\{\tau(t-[z_1^{-1}])
\tau(t-[z_2^{-1}])-\tau(t-[z_1^{-1}]-[z_2^{-1}])\tau(t)\}$$
expressed in terms of $\hw(t,z)$ instead of $\tau(t)$. This identity 
follows from Sato's bilinear identity.  $\Box$\\

Shifting the argument $t\mapsto t+[z_2^{-1}]$ and taking into account
(2.3) one gets $$(z_1-z_2)\{\hw(t+[z_2]^{-1},z_1)-\hw(t,z_1)\}
$$ $$=-\hw'(t+[z_2^{-1}],z_1)+\{w_1(t+[z_2^{-1}])-w_1(t)\}\hw(t+[z_2^{-1}],z_1)
.$$  For simplicity, let us introduce notations: $\hw$ is always $\hw(t,z_1)$, 
$\G$ will denote the shift $\G\hw=\hw(t+[z_2^{-1}],z_1)$. Thus, the statement 
of the lemma takes the form
$$(z_1-z_2)(\G\hw-\hw)=-\G\hw'+(\G w_1-w_1)\G\hw.$$This equation
contains all the equations of the hierarchy. They can be obtained by an
expansion of the equation in powers of $z_2^{-1}$ taking into account that
$\G=\exp(\sum_1^\infty\p_kz_2^{-k}/k)=\sum_0^\infty p_n^+\cdot z_2^{-n}$:
$$z_1\hw'=p_2^+\hw-\hw''+w'_1\hw, \eqno{(2.4)}$$
$$z_1p_2^+\hw=p_3^+\hw-p_2^+\hw'+w'_1\hw'+p_2^+w_1\cdot\hw, \eqno{(2.5)}$$
$$z_1p_3^+\hw=p_4^+\hw-p_3^+\hw'+p_3^+w_1\cdot\hw+p_2^+w_1\cdot\hw'+w'_1\cdot
p_2^+\hw\eqno{(2.6)}$$ and, generally,
$$z_1p_j^+\hw=p_{j+1}^+\hw-p_j^+\hw'+\sum_{l=0}^{j-1}p_{j-l}^+w_1\cdot p_l^+
\hw.\eqno{(2.7)}$$\\

{\bf 3. New identities.}\\

Let $$\G_n=\G-\sum_0^{n-1}p_i^+z_2^{-i}.$$ In other words, $\G_n$ is 
the operator $\G$ minus a part of its asymptotic in $z_2^{-1}$. Let
$$\bH=-\p+\G_1w_1+\hw\cdot\res_{z_1}.$$ The statement of the lemma of Sect.2 
can be presented as $$(z_1-z_2)\G_1\hw=-\G_1\hw'-\hw'+\G_1w_1\cdot\G_1\hw+\G_1
w_1\cdot\hw.$$ Taking into account that $\res_{z_1}\G_1\hw=\G_1w_1$, 
we find the following form of the Fay identity, 
$$(z_1-z_2)\G_1\hw=\bH(\G_1\hw)-\hw'.\eqno{(3.1)}$$

{\bf Proposition 1.} {\sl The equation $$(z_1-z_2)n^{(k)}=\bH(n^{(k)})-\p_k\hw
 \eqno{(3.2)}$$ has a unique solution $$n^{(k)}=\sum_1^\infty n^{(k)}_{j}z_2^{
-j},~~n^{(k)}_{j}=P_j^{(k)}\hw$$ where $P_j^{(k)}$ are differential operators 
in $\p_l$ having a form $\sum f^{\a_1,...,\a_s}p_{\a_1}^+...p_{\a_s}^+$ being
$\a_1,...,\a_s>0$, $s>0$. Coefficients $f$ do not depend on $z_1$.

In the case $k=1,2,$ and 3 the explicit expression for $n^{(k)}$ are}
$$n^{(1)}=\G_1\hw,\eqno{(3.3a)}$$ $$n^{(2)}=2z_2\G_2\hw-(\p-\G_1w_1)n_1,
\eqno{(3.3b)}$$ $$n^{(3)}=3z_2^2\G_3\hw-(\p-\G_1w_1)n_2-z_2(\p-\G_1w_1)\G_2\hw.
\eqno{(3.3c)}$$

{\em Proof}. The equation (3.2) is equivalent to a set of recurrence relations
$$n^{(k)}_{j}=z_1n^{(k)}_{j-1}+(n^{(k)}_{j-1})'-\sum_1^{j-2}p_m^+w_1\cdot n^
{(k)}_
{j-1-m}-n^{(k)}_{-1,j-1}\hw,~n^{(k)}_{1}=\p_k\hw \eqno{(3.4)}$$ where $n^{(k)}
_{-1,j-1}=\res_{z_1}n^{(k)}_{j-1}$ and the term with a sum is absent when $j-2
<1$. 
One can use induction. For $j=1$ the statement is true: one can prove a
formula $$\p_k=k\sum_{\a_1+...+\a_s=k}{(-1)^{s-1}\over s}p_{\a_1}^+...p_{\a_s}
^+$$ where the sum runs over all decompositions of $k$ into sums of
positive (all $\a_l>0$) integers\footnote{The proof is:
$\sum_1^\infty\p_kz_2^{-k}/k=\ln\exp\sum_1^\infty\p_kz_2^{-
k}/k=\ln\G=\ln(1+\G_1) =\sum_1^\infty(-1)^{s-1}\G_1^s/s$. Expanding in
powers of $z_2^{-1}$, one arrives at the required formula.}.
 
Let the statement be true for $n^{(k)}_{j-1}$. One can eliminate $z_1n^{(k)}_
{j-1}$ from Eq.(3.4) with the help of the Fay identity (2.7).
We find that $n^{(k)}_{j}$ has a needed form, however without guarantee that
there are no terms where all the $\a_l$'s vanish. If this were the case,
$n^{(k)}_{j}$ would contain terms of zero degree in $z_1^{-1}$. On the other
hand, expanding (3.4) in powers of $z_1^{-1}$ and taking the zero degree
term, one has: $n^{(k)}_{j}|_{z_1^0}=n^{(k)}_{-1,j-1}-n^{(k)}_{-1,j-1}=0$. This
proves the proposition for all the $j$'s and $k$'s. The uniqueness easily 
follows from the recurrence formula (3.4).

The additional statement (3.3) for  $k=1$ is just the Fay identity (3.1).
The others ($k=2$ and 3) can be obtained
by straightforward but cumbersome calculations, see Appendix. $\Box$\\

{\bf 4. The universal equation.}
Let $Q$ be a function. We call the equation $$f'+1-z_2f+Qf-\G_1w_1f=0
\eqno{(4.1)}$$ universal since it plays 
a fundamental role for all the constrained hierarchies, as we will see below.
This equation has exactly one solution in the form of a series
$f_Q=\sum_1^\infty f_jz_2^{-j}$. The coefficients $f_j$ are functions. It is
easy to see that $f_1=1$, $f_2=Q$, $f_3=Q'+Q^2-w'_1$ etc.\\

{\bf 5. $k$-constraint.}
The $k$-constraint is $L_-^k=q\p^{-1}r$ where $q$ and $r$
satisfy differential equations $$\p_nq=L_+^nq,~~\p_nr=-L_+^{n*}r\eqno{(5.1)}$$
and $L_+^{n*}$ is a formal adjoint operator. It is well-known 
that this constraint is compatible with the hierarchy equations. The equation
for $\hw(t,\p)$: $\p_n\hw(t,\p)=-L_-^n\hw(t,\p)$ transforms for $n=k$ to
$$\p_k\hw(t,\p)=-q\p^{-1}r\hw(t,\p).\eqno{(5.2)}$$

Dividing the last equation by $q$, and multiplying on the left by $\p$, one 
gets $$\p\circ\p_k(\hw(t,\p))-{q'\over q}\p_k(\hw(t,\p))=-qr\hw(t,\p).$$ The
parentheses in the l.h.s.
mean that the operator $\p_k$ acts on $\hw(t,\p)$,~ $q'=\p(q)$. Rewriting this
as $$\p_k(\hw(t,\p))\p+\p_k(\hw'(t,\p))-{q'\over q}\p_k\hw(t,\p)=-qr\hw(t,\p),
$$ one obtains for $\hw(t,z_1)$ the following:
$$z_1\p_k(\hw(t,z_1))+\p_k(\hw'(t,z_1))-{q'\over q}\p_k\hw(t,z_1)=-qr\hw(t,z_1
).$$Finally, let $Q=q'/q$ and $R=qr$, then $$\p_k(\hw'(t,z_1))+(z_1-Q)\p_k(
\hw(t,z_1))+R\hw(t,z_1)=0. \eqno(5.3)$$ The first term of expansion in
powers of $z_1^{-1}$ yields $R=-\p_kw_1$.\\

{\bf 6. Conditions that $\hw$ belongs to a constrained hierarchy.} Suppose one
wants to find necessary and sufficient conditions that
a Baker function $\psi(t,z_1)=\hw(t,z_1)\exp\xi(t,z_1)$ of the KP
hierarchy belongs to the $k$-constrained subhierarchy. (With some abuse of 
terminology, we also call $\hw$ a Baker function). This problem can be solved
easily. If $\hw$ belongs to the constrained hierarchy then Eq.(5.3) holds and
$${\p_k\hw'+z_1\p_k\hw-\p_kw_1\cdot\hw\over\p_k\hw}=Q,\eqno{(6.1)}$$ and
therefore the left-hand side of this equation is independent of $z_1$. 
Conversely, let the expression {(6.1)} where $\hw$ is a Baker function of the 
KP hierarchy not depend on $z_1$. Then letting $R=-\p_kw_1$ and solving the
equations $Q=q'/q$ and $R=qr$ with respect to $q$ and $r$, one easily obtains 
(5.2). It is known (see [2]) that this implies that $q$ and $r$ satisfy
Eq.(5.1), i.e., $\hw$ belongs to the $k$-constrained hierarchy. Thus, 
the following criterion holds:\\

{\bf Proposition 2.} {\sl A Baker function of the KP hierarchy belongs to 
the $k$-constrained hierarchy if and only if the expression (6.1) is 
independent of $z_1$}.\\

The term with $z_1$ in (6.1) can be eliminated with the aid of (2.4-7). E.g., for
$k=1$ using (2.4) we have $$\p_1\hw'+z_1\p_1\hw-w'_1\hw=\hw''+z_1\hw'-w'_1\hw'
=p_2^+\hw$$ and the condition reads: $p_2^+\hw/\hw'$ must be independent of
$z_1$. The same also can be expressed as $$\left|\begin{array}{cc}w'_{i_1}&w'_
{i_2}\\p_2^+w_{i_1}&p_2^+w_{i_2}\end{array}\right|=0\eqno{(6.2)}$$where $i_1$
and $i_2$ are arbitrary.

In the case $k=2$ we have $$\p_3\hw-z_1\p_2\hw=p_3^+
\hw-w'_1\hw'-\p_2w_1\cdot\hw$$ Then the expression (6.1) takes the form, using
(2.5),$$Q={\p_2\hw'+\p_3\hw-p_3^+\hw+w'_1\hw'\over\p_2\hw}={
2p_3^+\hw-p_2^+\hw'+w'_1\hw'\over\p_2\hw}.$$We also have used here the 
known explicit expressions for Schur polynomials: $$p_2^+=
\frac 12(\p^2+\p_2),~~p_3^+=\frac 13\p_3+p_2^+\p-\frac 13\p^3.$$ 
Independence of $z_1$ of the expression $Q$ is equivalent 
to $$\left|\begin{array}{cc} \p_2w_{i_1}&\p_2w_{i_2}\\(\p_3+\p_2\p-p_3^++w'_1\p
)w_{i_1}&(\p_3+\p_2\p-p_3^+
+w'_1\p)w_{i_2}\end{array}\right|=0$$or$$\left|\begin{array}{cc}
(2p_2^+-\p^2)w_{i_1}&(2p_2^+-\p^2)w_{i_2}\\(2p_3^+\hw-p_2^+\hw'+w'_1\hw')w_{i_1
}&(2p_3^+\hw-p_2^+\hw'+w'_1\hw')w_{i_2}\end{array}\right|=0.\eqno{(6.3)}
$$ For $k=3$, as it can be shown, the condition is

$$\left|\begin{array}{cc}P_1^{(3)}w_{i_1}&P_1^{(3)}w_{i_2}\\P_2^{(3)}w_{i_1}&
P_2^{(3)}w_{i_2}\end{array}\right|=0\eqno{(6.4)}$$ where 
$$P_1^{(3)}=3p_3^+-3p_2^+\p+\p^3,~~P_2^{(3)}=3p_4^+-3p_3^+
\p+p_2^+\p^2+3w'_1p_2^+-2w'_1\p^2-w''_1\p.$$
All this is simple and straightforward and would not deserve a special
discussion if not the following generalization. As it will be shown in the next
sections, infinite sequences of
determinant identities can be written in each of cases $k=1$, $k=2$ and $k=3$
such that (6.2), (6.3) and (6.4) are but first terms of them. A similar
statement is also true for any $k$.\\

{\bf 7. Main results about constrained hierarchies.} Here we make a statement
and derive some corollaries of it. In the next section we give a proof.\\

{\bf Proposition 3.} {\sl If $n^{(k)}$ are those defined in Sect.3 then
for a $k$-constrained hierarchy 
we have $${n^{(k)}\over\p_k\hw}=f_Q, ~k=1,2,3,...\eqno{(7.1)}$$
Here $f_Q$ is the solution of the universal equation, Sect.4, and $Q=q'/q$.
Hence the
left-hand side of the equation (7.1) does not depend on $z_1$.}\\

What we have done actually, is that we constructed some expressions having the
asymptotic $z_2^{-1}+Qz_2^{-2}+...$
with a property: if the first nontrivial term $Q$ does not depend on $z_1$ then
neither does the whole series.

The numerators of these expressions 
can be expanded in $z_2^{-1}$: $$ n^{(k)}=\sum_{j=1}^\infty P_j^{(k)}\hw\cdot
z_2^{-j}$$ where
$P_j^{(k)}$ are differential operators. Namely,
$$P_j^{(1)}=p_j^+,~P_j^{(2)}=\sum_{l=1}^{j-1}p_l^+w_1\cdot
p_{j-l}^+-p_{j}^+\p+2p_{j+1}^+$$ and
$$P_j^{(3)}=\sum_{\a,\b,\g>0,\a+\b+\g=j}p_\a^+w_1\cdot p_\b^+w_1\cdot
p_\g^+-\sum_{\a=1}^{j-1}(2p_\a^+
w_1\cdot p_{j-\a}^+\p+p_\a^+ w'_1\cdot p_{j-\a}^+)+3\sum_1^{j-1} p_\a^+
w_1\cdot p_{j+1-\a}^+$$
$$+p_{j}^+\p^2-3p_{j+1}^+\p+3p_{j+2}^+.$$

{\bf Corollary 4.} {\sl The equalities 
$$P_j^{(k)}w_i=f_j\cdot \p_kw_i,~~j\geq 1$$ hold.}\\

{\bf Corollary 5.} {\sl There are identities 
$$\left|\begin{array}{cc}P_{j_1}^{(k)}w_{i_1}&P_{j_1}^{(k)}w_{i_2}\\P_{j_2}^{(k
)}w_{i_1}&P_{j_2}^{(k)}w_{i_2}\end{array}\right|=0$$ where $i_1$,
$i_2$, $j_1$ and $j_2$ are arbitrary $\geq 1$.}\\

The equalities (6.2-4) are the first of these identities, with the least 
possible $j_1$ and $j_2$.\\

{\bf 8. Proof of the proposition 3.} First we prove that the left-hand side of
Eq.(7.1) satisfies a slightly
different equation $$f'=-1+z_2f+\G_1w_1\cdot f-Qf-{\p_kw_1\cdot
\hw\over\p_k\hw}(f-f_0)\eqno{(8.1)}$$ where $f_0$ is
the limiting value of $f$ when $z_1\rarr \infty$: $f_0=n^{(k)}_{-1}/\p_kw_1$, 
where $n^{(k)}_{-1}=\res_{z_1}n^{(k)}$. If we manage to prove this,
then, passing to the limit $z_1\rarr\infty$ in the equation (8.1), we get for 
$f_0$ :$$f'_0=-1+z_2f_0+\G_1w_1f_0-Qf_0.$$ Subtracting this equation from 
Eq.(8.1), we obtain for $g=f-f_0$:
$$g'=z_2g+\G_1w_1\cdot g-Qg-{\p_kw_1\cdot \hw\over\p_k\hw}g.$$ This is a 
homogeneous equation, $g$ is a double series in $z_1^{-1}$ and $z_2^{-1}$. Let 
$g=\sum_{j^*}^\infty g_jz_2^{-j}$ where $g_{j^*}$ is the
first non-zero coefficient assuming that it exists. Then expanding the equation 
for $g$ in powers of $z_2^{-1}$ one gets for the coefficient of $z_2^{-j^*+1}$: 
$g_{j^*}=0$ in contradiction to the assumption.
This means that $g\equiv 0$. Then $f=f_0$, $f$ does not depend on $z_1$ and
satisfies the universal equation,
as required. Thus, it suffices to prove that the left-hand side of Eq.(7.1),
which will be denoted as $f$, satisfies Eq.(8.1).

We have $f=n^{(k)}/\p_k\hw$. What equation for $n^{(k)}$ does follow from 
Eq.(7.1)? Firstly, $$f'={(n^{(k)})'\over\p_k\hw}-f\cdot{\p_k\hw'\over\p_k\hw}.
$$ 
Eliminating $\p_k\hw'$ with the help of the constraint equation (5.3) one 
obtains
$$f'={(n^{(k)})'\over\p_k\hw}+(z_1-Q)f-{\p_kw_1\cdot\hw\over\p_k\hw}f.$$ 
Substituting
this expression for $f'$ in (8.1) one gets the equation for numerators
$$(z_1-z_2)n^{(k)}=-(n^{(k)})'+\G_1w_1\cdot n_k+n^{(k)}_{-1}\hw-\p_k\hw.\eqno
{(8.2)}$$ 
Here $n^{(k)}_{-1}=f_0\cdot\p_kw_1=\res_{z_1}n^{(k)}$. This equation is nothing 
less than
the equation of the proposition 1, and $n^{(k)}$'s are its solutions. $\Box$\\

{\bf Acknowledgement.} We are thankful to the {\em Mathematisches 
Forschungsinstitut
Oberwolfach} whose {\em RiP} program helped us to accomplish this work.\\

\vspace{.5in}

\centerline{{\bf Appendix. Proof of the Proposition 1 for $k=2$ and 3.}}

\vspace{.1in}
Recall the definition of the operator $\bH$:
$$
\bH(v)=-(\partial-\G_1w_1)v+v_{-1}\hat{w}=-v'+\G_1w_1\cdot v+v_{-1}\hat{w}\,,$$
here $$v_{-1}=\res_{z_1}(v)\,.$$
Thus, for instance,
\begin{eqnarray*}
n^{(1)}_{-1}&=&\G_1w_1\,,\\
n^{(2)}_{-1}&=&2 z_2 \G_2w_1-(\partial-\G_1w_1)n^{(1)}_{-1}\,,\\
n^{(3)}_{-1}&=&3 z_2^2 \G_3w_1-(\partial-\G_1w_1)n^{(2)}_{-1}
        -z_2(\partial-\G_1w_1)\G_2w_1\,,
\end{eqnarray*}
We prepare a few properties of $\bH$. For any function $v$
$$\bH(\G_1w_1\cdot v)=\G_1w_1\cdot \bH(v)-\G_1w_1'\cdot v$$ and $$
-(\partial-\G_1w_1)\bH(v)=\bH(-(\partial-\G_1w_1)v)-v_{-1}\hat{w}'\,.
$$

The following lemmata will be frequently needed.\\

{\bf Lemma 1.} {\sl Let 
\[(z_1-z_2)v=\bH(v)-u\] then} 
\[(z_1-z_2)(-(\partial-\G_1w_1)v)=\bH(-(\partial-\G_1w_1)v)-v_{-1}\hat{w}'
                               +(\partial-\G_1w_1)u\,.\]
This is obvious.\\

{\bf Lemma 2.} {\sl The following holds:}
\begin{eqnarray}
(z_1-z_2)\G_1\hat{w}&=&\bH(\G_1\hat{w})-p_1^+\hat{w}\,,\\
(z_1-z_2)z_2\G_2\hat{w}&=&\bH(z_2\G_2\hat{w})+(-p_2^+\hat{w}
                        +\G_1w_1\cdot p_1^+\hat{w})\,,\\
(z_1-z_2)z_2^2\G_3\hat{w}&=&\bH(z_2^2\G_3\hat{w})+z_2(\G_1w_1\cdot p_1^+
\hat{w})                       \nonumber\\
                   & &
+(-p_3^+\hat{w}+\G_1w_1\cdot p_2^+\hat{w}-w'_1p_1^+\hat{w})\,,
\end{eqnarray}

{\em Proof.} One must replace $\G_2\hw$ by $\G_1\hw-z_2^{-1}\hw'$
and $\G_3\hw$ by $\G_1\hw-z_2^{-1}\hw'-z_2^{-2}p_2^+\hw$, then 
apply equations (3.1) and (2.4-6). Finally, one must go back to
$\G_2\hw$ and $\G_3\hw$. The calulations are not difficult.
$\Box$ 

Now we are proving the proposition. In the case $k=1$ there is 
nothing to prove, the required equation coincides with the Fay 
identity (3.1). 

In the case $k=2$ Lemma 2 yields
$$(z_1-z_2)2z_2\G_2\hat{w}=2\bH(z_2\G_2\hat{w})+2(-p_2^+\hat{w}
+2\G_1w_1\cdot p_1^+\hat{w}).$$ Lemma 1, starting from (3.1),
implies
\begin{eqnarray*}
(z_1-z_2)(-(\partial-\G_1w_1)n^{(1)})&=&\bH(-(\partial-\G_1w_1)n^{(1)})\\
  & &-n^{(1)}_{-1}\hat{w}'+(\partial-\G_1w_1)\hat{w}'\\
&=&\bH(-(\partial-\G_1w_1)n^{(1)})-2\G_1w_1\hat{w}'+\hat{w}''\,.
\end{eqnarray*}
A sum of two last equations is exactly what was to prove.

Now, let $k=3$. Applying Lemma 2 again, we have
\begin{eqnarray*}
(z_1-z_2)3z_2^2\G_3\hat{w}&=&\bH(3z_2^2\G_3\hat{w})+3z_2(\G_1w_1\cdot p_1^+\hat
{w})\\                   & &
+3(-p_3^+\hat{w}+\G_1w_1\cdot p_2^+\hat{w}-w'_1p_1^+\hat{w})\,,
\end{eqnarray*}
while Lemma 1 and the equation we have just proven for $k=2$ yield:
\begin{eqnarray*}
(z_1-z_2)(-(\partial-\G_1w_1)n^{(2)})&=&\bH(-(\p-\G_1w_1)n^{(2)})\\
                    & &-n^{(2)}_{-1}\hat{w}'+(\partial-\G_1w_1)\partial_2\hat{w}
\,.
\end{eqnarray*}
Finally we use the equation for $\G_2\hat{w}$ and apply Lemma 1 to it:
\begin{eqnarray*}
(z_1-z_2)z_2(-(\partial-\G_1w_1)\G_2\hat{w})&=&
H(z_2(-(\partial-\G_1w_1)\G_2\hat{w}))\\
                    & &-z_2\G_2w_1\cdot \hat{w}'\\
                    & &+(\partial-\G_1w_1)(p_2^+\hat{w}-\G_1w_1\cdot \hat{w}') 
\,.
\end{eqnarray*}
On addition of the last three equations we obtain:
\begin{eqnarray*}
(z_1-z_2)n^{(3)}&=&\bH(n^{(3)})\\
                 & &-3w_1'\hat{w}'
                    +3z_2(\G_1w_1\cdot \hat{w}'-\G_2w_1\cdot \hat{w}')\\ & &-
3p_3^+\hat{w}+\partial\partial_2\hat{w}
                    +\partial p_2^+\hat{w}\\
                 & &+2\G_1w_1\cdot p_2^+\hat{w}
                    -\G_1w_1\cdot \partial_2\hat{w}
                    -\G_1w_1\cdot \hat{w}''\,,
\end{eqnarray*}
which gives the claim by using obvious identities for
the Schur polynomials $p_2$ and $p_3$ and the equality
$\G_1w_1-\G_2w_1=z_2^{-1}p_1^+w_1$.\\

{\bf References.}\\

\noindent {\bf 1.} Y. Cheng, W. Strampp, B. Zhang, {\em CMP}, {\bf 168}
(1995), 117-135.\\

\noindent {\bf 2.} H. Aratyn, Integrable Lax hierarchies, their 
symmetry reductions and multi-matrix models, Preprint, hep-th 9503211,
1995\\

\noindent {\bf 3.} P. Van Moerbeke, Integrable foundations of string 
theory, in: {\em Lectures on Integrable Systems} in Memory of Jean-Louis 
Verdier, eds. Babilon, Cartier and Kosmann-Schwarzbach, {\em World      
Scientific} (1994) 163-268. 
\end{document}